\documentclass[aps,pra,amsmath,amssymb,showpacs, twocolumn]{revtex4}

\usepackage[T1]{fontenc}
\usepackage{amssymb,amsmath}
\usepackage{amsfonts}
\usepackage{graphicx, pstricks,epsfig}

\usepackage{graphicx}

\def\ket#1{|#1\rangle}
\def\bra#1{\langle#1|}

\def\idmat{\mathbf{1}}
\def\caln{\mathcal{N}}
\def\calc{\mathcal{C}}

\def\tr{\mathrm{tr}}
\def\bfu{\mathbf{u}}

\def\bfmu{\mbox{\boldmath$\mu$}}

\begin{document}

\title{Parametrization of spin-1 classical states}
\author{Olivier Giraud$^{1}$, Petr Braun$^{2,3}$ and Daniel Braun$^{4}$}
\affiliation{$^{1}$ Univ.~Paris-Sud, CNRS, LPTMS, UMR 8626, Orsay, F-91405, France\\
$^{2}$ Fachbereich Physik, Universit\"at Duisburg--Essen, 47048 Duisburg,
  Germany\\
$^{3}$ Institute of Physics, Saint-Petersburg University, 198504
Saint-Petersburg,  Russia\\
 $^{4}$ Laboratoire de Physique
Th\'eorique, Universit\'e de Toulouse, CNRS,  31062 Toulouse,
France}
\begin{abstract}
We give an explicit parametrization of the set of mixed quantum
states and of the set of mixed classical states for a spin--1. Classical
states are defined as states with a positive Glauber-Sudarshan
P-function. They are at the same time the separable symmetric states of two
qubits.  We explore the
geometry of this set, and show that its boundary 
consists of a two-parameter family of ellipsoids.  The boundary does not
contain any facets, but includes straight-lines corresponding to mixtures of
pure classical states. 
\end{abstract}

\date{September 30, 2011}

\maketitle

\section{Introduction}
The rise of quantum information theory has led to a large interest in
the geometry of specific sets of quantum states \cite{Bengtsson06}. The
most general
quantum state of a quantum system with $d$-dimensional Hilbert space
${\cal H}$ is given by a density operator $\rho$ that acts on ${\cal
  H}$. The density operator is a Hermitian, semi-definite positive
operator with trace
1. Diagonalization shows immediately that it can always be written as
a convex sum of projectors onto its eigenstates.  The set ${\cal N}$ of all
physical density operators is therefore the convex hull of projectors
onto all pure states in ${\cal H}$.
Certainly the most popular set of states in quantum information theory
is the  set of separable states, defined for a physical system that
can be partitioned into at least two subsystems.
If $\rho$ can be written as a convex sum of tensor products of
projectors onto pure states of the subsystems, that state is called
``separable'', and ``entangled'' otherwise \cite{Werner89}.  Clearly,
the set ${\cal S}$ of all separable states is a subset of ${\cal N}$.
Knowing the geometry, and in particular the surface $\partial{\cal S}$
of the set ${\cal
  S}$, is an important but difficult problem, as it would allow to
determine immediately whether a given state is inside or outside
${\cal S}$, or, in other words, whether it is entangled or not.  In the case
of two qubits, $\partial{\cal S}$ was shown to be smooth in the
interior of ${\cal N}$ \cite{Djokovic06}. Furthermore, for a general
bipartite state, it was shown that $\partial{\cal S}$ is not a
polytope \cite{Ioannou06} and, using non-linear entanglement
witnesses, that it does not contain any
facets \cite{Guehne07}.

Recently we introduced the convex set ${\cal C}\subseteq{\cal N}$ of
``classical states'' of a spin (or angular momentum) with total angular momentum
$j$ \cite{Giraud08}.    It is defined as the convex hull
of projectors
onto coherent states of $SU(2)$, which have the physical interpretation of
having minimal quantum uncertainty of the angular momentum vector, i.e.~they
resemble as much as possible a point in classical phase space.  The
interest of classical states is that they are defined even for a
single spin, i.e.~when the question of entanglement does not
even arise. Furthermore, they allow a definition of what a genuinely ''quantum''
state might be.  Indeed, one may define a measure of ``quantumness''
\cite{Giraud10} of a spin state
by measuring its distance from ${\cal C}$, just as the distance to
${\cal S}$ provides a measure of entanglement (see \cite{Horodecki09}
for an overview of this type of entanglement measure).    If distance
from ${\cal C}$ is measured through the Bures distance
\cite{Miszczak09,Bengtsson06}, quantumness of symmetric multi-qubit
states becomes essentially equivalent to their geometrical
entanglement \cite{Martin10}. Also note that the set of classical states of
a spin--1 is identical to the set of separable symmetric states (under the
exchange of particles) of two qubits. 

States of a spin with maximal quantumness with given total angular momentum
$j$ (i.e.~with 
Hilbert space dimension $2j+1$), the ``Queens
of Quantum'', can always
be found among pure states \cite{Giraud10}.
However, if $\partial{\cal C}$  contains facets, there might exist mixed
states with the same 
maximal quantumness. Knowing the form of the surface of the set of
classical states is therefore important. In \cite{Giraud10} it was shown
that for spin--$1$ states 
maximal quantumness is reached only for pure states, but for larger
$j$ maximally quantum states might comprise mixed states.  After what
was said above about
how little is known about the surface $\partial{\cal S}$  of the set of separable states, one
might expect that determining $\partial{\cal C}$ is a difficult
problem as well.  This is indeed the case, but nevertheless, here we
give a complete characterization of $\partial{\cal C}$ for the case of
a qutrit (i.e.~a three state system, corresponding to a pseudo-angular
momentum $j=1$).  We show that in this case ${\cal C}$, as ${\cal S}$,  is not a
polytope either, but rather a continuous family of
ellipsoids. We also show that the surface of ${\cal C}$ contains
families of straight lines.

\section{Spin--$\frac12$ case}
Let us first consider the trivial case of a spin--$\frac12$
system. In this case it was shown in  \cite{Giraud08} that the set
$\calc$ of classical states coincides with $\caln$. Any $2\times
2$ density matrix can be expanded over the basis of Pauli matrices
$\sigma_a$ as
\begin{equation}
\rho =\frac{1}{2}\idmat_{2}+\sum_a u_a\sigma_a
\label{canonrhoj12}
\end{equation}
with $\idmat_{2}$ the $2\times 2$ identity matrix and $u_a$ are real numbers with $a=x,y,z$. The matrix
$\rho$ given by \eqref{canonrhoj12} is Hermitian and has trace 1,
therefore it belongs to $\caln$
if and only if it is positive. The characteristic polynomial
of $\rho$ can be put under the form
\begin{equation}
\det\left(x \idmat_2-\rho\right)=x^2-x+\frac{1-\sum_au_a^2}{4}.
\end{equation}
Its roots are positive if and only if
\begin{equation}
\label{sphereC12}
\sum_au_a^2\leq 1.
\end{equation}
Equation~\eqref{sphereC12} is thus the necessary and sufficient
condition for $\rho\in\caln$ in terms of the coordinates $u_a$
which parametrize $\rho$. The boundary $\partial\caln$ of $\caln$
corresponds to points where one of the eigenvalues of $\rho$
vanishes. In terms of coordinates $u_a$ it is given by the
equation $\sum_au_a^2=1$. The parametrization of $\partial\caln$
is the parametrization of a sphere.

The results above correspond to the usual
picture of the Bloch sphere for spin--$\frac12$. The vector $\mathbf{u}$ is the Bloch
vector, and the boundary of (classical) states is the boundary of
the sphere, corresponding to rank-one matrices, or pure states,
with Bloch vector of length 1. Such a simple picture does not
exist for higher spins. Let us now consider the case of spin--$1$
states.

\section{Classicality criterion for spin--1 states}
We start with the expansion of a mixed spin--1 state over the
basis formed by the $3\times 3$ angular momentum matrices,
$J_{a}$, $a=x,y,z$, together with the $(J_{a}J_{b}+J_{b}J_{a})/2 $
and the $3\times 3$ identity matrix $\idmat_3$. We define a vector
${\bf u}$ and a matrix $W$ through coefficients of this expansion,
as
\begin{equation}
\rho =\frac{1}{3}\idmat_{3} +\frac{1}{2}\mathbf{u}.\mathbf{J}+\frac{1}{2}\sum_{a,b=x,y,z}\left(
W_{ab}-\frac{1}{3}\delta _{ab}\right)
\frac{J_{a}J_{b}+J_{b}J_{a}}{2}.
\label{canonrhoj1}
\end{equation}
The coefficients ${\bf u}$ and $W$ are related with $\rho$ through
\begin{equation}\label{uwviarho}
u_{a}=\tr\left( \rho J_{a}\right) ,\quad W_{ab}=\tr\rho \left(
J_{a}J_{b}+J_{b}J_{a}\right) -\delta _{ab}.
\end{equation}
Note that ${\bf u}$ is real and that $W$ is a real symmetric
matrix, with $\tr\ W=1$.

The expression \eqref{canonrhoj1} ensures that $\rho$ is Hermitian
with $\tr\rho=1$. Thus the set $\caln$ of density matrices is
the set of matrices of the form  \eqref{canonrhoj1} with $\rho\geq 0$.
According to \cite{Giraud08}, $\rho$ is a density matrix
associated with a classical state if and only if the real
symmetric $3\times 3$ matrix $Z$ with matrix elements
\begin{equation}
\label{condC}
Z_{ab}=W_{ab}-u_a u_b
\end{equation}
is non-negative, thus the set $\calc$ of classical
density matrices is the set of matrices of the form  \eqref{canonrhoj1} with $Z\geq 0$.

\section{Set $\caln$ of density matrices}
\label{paramN}

The class of density matrices $\caln$ comprises Hermitian
non-negative matrices with trace 1. Its parametrization is
important in many applications and can be achieved in several
ways. One of these is based on the representation $\rho=U\;
\mathrm{diag}[\lambda_1..\lambda_{2j+1}]\;U^{-1},\quad
\lambda_i\ge 0,\quad \sum\lambda_i=1$, where $U$ runs over a subset of
the unitary group chosen such that each $\rho$ is obtained once
and only once. A parametrization for the case $j=1$ using Gell-Mann
matrices   is
considered in \cite{Byrd01}; see also \cite{Dennis04b}  for the
closely related problem of $3\times 3$ coherence matrices of
nonparaxial light. Another method uses the factorization $\rho=V
V^\dagger$ where $V$ is upper triangular \cite {Chung75}. Here we shall
give an alternative representation based on the formula
(\ref{canonrhoj1}).

Parameters  ${\bf u}$  and  $W$ have the nice feature, similar to
the Bloch picture in the two-dimensional case, that under rotation
of the coordinate system with an orthogonal rotation matrix $O$,
$\rho$ is transformed into a matrix with parameters $O{\bf u}$ and
$O WO^T$, i.e.~${\bf u}$ and $W$ transform with the same
rotation $O$. Thus it will be convenient to express them in a basis where
$W$ is diagonal, $W=\mathrm{diag}[\mu_x,\mu_y,\mu_z]$; we shall
write the result as
\begin{eqnarray}
\rho=\rho'+\frac{1}{2}\mathbf{u}.\mathbf{J},\nonumber\\
\label{reducedrho}
\rho'=\frac{1}{2}\left(\mu_x J_x^2+\mu_y
J_y^2+\mu_z J_z^2\right) .
\end{eqnarray}
Considering that $\tr\; W=1$ and that, in a state $\rho$ with
angular momentum 1, we have $0\le\tr\; \rho J_a^2\le 1,\quad 0\le
|\tr \rho \mathbf{J}|\le 1$   we obtain the necessary conditions on the
parameters in (\ref{reducedrho}),
\begin{eqnarray}
\sum_{a=x,y,z}\mu_a=1,\nonumber\\
-1\le \mu_a\le 1,\quad a=x,y,z,\label{limitonmu}\\
u_x^2+u_u^2+u_z^2\le 1\label{limit0onu}.
\end{eqnarray}

For the ``truncated''  matrix $\rho'=\left
.\rho\right|_{\mathbf{u}= 0}$ conditions \eqref{limitonmu} are
also sufficient to guarantee that  $\rho'\in \caln$. Indeed,
direct calculation shows that the eigenvalues of $\rho'$
 are
\begin{eqnarray}
\lambda_a'=\frac{1-\mu_a}2\ge 0,\quad a=x,y,z,
\end{eqnarray}
while the corresponding eigenvectors $\ket{v_a}$  are eigenvectors of $J_a$
with eigenvalue zero. Since $\langle
v_a|\mathbf{J}|v_a\rangle=\mathbf{0}$, we have $\langle
v_a|\rho|v_a\rangle=\lambda_a'$. These averages give an upper
bound to the smallest eigenvalue of $\rho$. It immediately follows
that if $\rho$ belongs to $\caln$ then so does $\rho'$ but not
vice versa.

In fact, a stronger statement can be made. Let
$\rho_\kappa=\rho'+\frac{\kappa}{2} \mathbf{u.J}$ be a density
matrix differing from $\rho$ by a positive factor
$\kappa$ in the part linear in $\mathbf J$. Then the lowest 
eigenvalue of $\rho_\kappa$ is a monotonically decreasing function
 of $\kappa$.  Consequently if
$\rho_\kappa$ with some $\kappa=\kappa_1$ belongs to $\caln$ then
 so do all matrices with  $0\le\kappa<\kappa_1$. These assertions
follow from the following theorem of perturbation theory (for a
proof, see the Appendix): Let $H=H_0+\kappa V, \;\kappa\geq 0,$ be
a Hermitian matrix whose spectrum is bounded from below, and
$E_0(\kappa),\psi_0(\kappa)$ be its lowest eigenvalue and
eigenstate. Suppose that $E_0(0)$ is non-degenerate
 and $\langle\psi_0(0)|V|\psi_0(0)\rangle=0$.  Then
$E_0(\kappa)$ is a monotonically decreasing function. Setting
$H_0\to \rho',\;V\to (1/2)\mathbf{u.J}$ we come to the statement
above.

Let us find the constraints sufficient and necessary to guarantee
non-negativity of $\rho$. The characteristic polynomial of $\rho$
written in the form (\ref{canonrhoj1}) can be presented as
\begin{equation}
\det\left(x \idmat_3-\rho\right)=x^3-x^2+a x-\det\rho
\end{equation}
with
\begin{equation}
a=\frac{1}{4}\left(-|\bfu|^2+1-\frac{\tr W^2-1}{2}\right).
\end{equation}
Since $\rho$ is Hermitian the three roots of the polynomial are
real. According to Descartes' rule of signs, a polynomial of the
form $x^3-x^2+ax-b$ with three real roots has all its roots
positive if and only if $a$ and $b$ are positive. Thus
$\rho\in\caln$ iff $\det\rho\geq 0$ and
\begin{equation}
\label{condN1}
1-|\bfu|^2+\frac{1-\tr W^2}{2}\geq 0.
\end{equation}
The latter condition defines a sphere in the $u$-space. Since it does not depend on the basis in which
$\bfu$  and $W$ is expressed, we can write Eq.~\eqref{condN1} in the basis
where $W$ is diagonal; in that basis it becomes
\begin{equation}
\label{sphereN} |\mathbf{u}|^2\leq
1+\mu_x\mu_y+\mu_x\mu_z+\mu_y\mu_z.
\end{equation}
 One can
check that the condition $\det\rho\geq 0$ can be rewritten
\begin{equation}
\label{condN2}
\langle \bfu|W|\bfu\rangle-|\bfu|^2+\frac{1-\tr W^2}{2}-\det{W}\geq 0,
\end{equation}
which in the basis where $W$ is diagonal becomes
\begin{equation}
\label{ellipsoidN} \sum_au_a^2(1-\mu_a)\leq
(1-\mu_x)(1-\mu_y)(1-\mu_z).
\end{equation}
When all $\mu_a$ differ from 1, it defines an
ellipsoid in the $u-$space lying inside
 both spheres \eqref{limit0onu} and \eqref{sphereN}. Indeed, the squared
radius of the ellipsoid along the $x$-axis for instance is given by
\begin{equation}
 r_x^2=(1-\mu_y)(1-\mu_z)=\mu_x+\mu_y\mu_z
\end{equation}
and using the fact that $1-\mu_x^2\geq 0$ we get
\begin{equation}
\label{rx1}
 r_x^2\leq 1-\mu_x^2+r_x^2=1+\mu_x\mu_y+\mu_x\mu_z+\mu_y\mu_z.
\end{equation}
On the other hand, the inequality $\mu_y\geq -\mu_z$ (coming from $\mu_x\leq 1$)
yields
\begin{equation}
\label{rx2}
 r_x^2=(1-\mu_y)(1-\mu_z)\leq 1-\mu_z^2\leq 1,
\end{equation}
thus the ellipsoid
also lies within the sphere of radius 1.  However, when  one or
two $\mu_a$ are equal to 1, then Eqs.~\eqref{limit0onu} and
\eqref{sphereN} have to be taken into account. We
finally obtain that $\rho\in\caln$ if and only if the $(u_a,\mu_a)$ verify one of the
following conditions:
\begin{enumerate}
\item All $\mu_a$ are such that $-1\leq \mu_a<1$, and
\begin{equation}
\label{ensembleN}
\sum_a \frac{u_a^2}{\mu_a+\mu_b\mu_c}\leq 1,
\end{equation}
with $\sum_a\mu_a=1$ and $b,c$ are the two indices which differ
from $a$ (this automatically implies  \eqref{limit0onu},\eqref{sphereN});
\item Exactly one $\mu_a$ is equal to 1, say $\mu_z=1$. Then
$\mu_y=-\mu_x$ with $-1<\mu_x<1$, and
$u_x=u_y=0$. Equation \eqref{sphereN} yields  $u_z^2\leq 1-\mu_x^2$ and is obviously more restrictive than \eqref{limit0onu};
\item Two of the $\mu_a$ are equal to 1,  say $\mu_y=\mu_z=1$, then
$\mu_x=-1$, $u_x=u_y=u_z=0$. This corresponds, up to rotation, to the state
$\ket{1,0}\bra{1,0}$ (in $|j,m\rangle$ notation).
\end{enumerate}

The boundary $\partial\caln$ of $\caln$ corresponds to points
where one of the inequalities \eqref{condN1} or \eqref{condN2}
becomes an equality.  For $\mu_a\neq 1$ (case 1 above), Eq.~\eqref{condN2},  equivalent to the equation of the ellipsoid
\eqref{ellipsoidN},  is more restrictive than  \eqref{condN1} 
 so that  $\partial\caln$ coincides with the  surface of the
ellipsoid \eqref{ensembleN}. The cases when one or two $\mu_a$ are equal to 1 correspond to
 cases 2 and 3, where equality is reached in \eqref{ellipsoidN}.
Therefore, the surface  $\partial\caln$ is the union of points corresponding to case 1
when \eqref{ensembleN} is an equality, and of points corresponding to cases 2 and 3.

Points of $\partial\caln$ corresponding to case 1, with equality
in \eqref{ensembleN}, belong
to a two-parameter set of ellipsoids that can be
parametrized by $\mu_a\in[-1,1[$ and
\begin{equation}
\label{parametrizationN}
\bfu=\begin{pmatrix}
\sqrt{\mu_x+\mu_y\mu_z}\cos\theta\cos\varphi \\
\sqrt{\mu_y+\mu_x\mu_z}\cos\theta\sin\varphi \\
\sqrt{\mu_z+\mu_x\mu_y}\sin\theta\ \ \ \ \ \ \
\end{pmatrix},\ \ \ \ \theta\in[0,\pi],\ \ \varphi\in[0,2\pi[.
\end{equation}

Any density matrix $\rho\in\caln$ for spin 1 is parametrized by 8
real numbers. Thus $\partial\caln$ should be parametrized by 7
numbers. For points corresponding to case 1,
besides $\mu_1$, $\mu_2$, $\theta$ and $\varphi$, the three
remaining parameters correspond to the three angles that
parametrize the orthogonal matrix required to diagonalize $W$. These
orthogonal transformations also include transpositions of axes
$x,y,z$; to get each matrix once and only once we must introduce
restrictions on $\mu_a$, say, $\mu_z\le\mu_y\le\mu_x$. One
way to do so is to use the eigenvalues $\lambda_a'$ of $\rho'$ as auxiliary
variables, with $\mu_a=1-2\lambda_a'$, setting
\begin{eqnarray}
\label{parametrizationNmu}
 \begin{pmatrix}
\lambda_x' \\
\lambda_y' \\
\lambda_z'
\end{pmatrix}=
 \begin{pmatrix}
\sin^2\theta'\sin^2\varphi' \\
\sin^2\theta'\cos^2\varphi' \\
\cos^2\theta'
\end{pmatrix}
\end{eqnarray}
with $\theta'\in ]0,\arctan (1/\cos( \varphi'))]$ and $\varphi'\in]0,\pi/4]$.
Points of $\partial\caln$ corresponding to cases 2 and 3
are of measure zero on the surface.

\section{Set $\calc$ of classical state density matrices}
\label{paramC} We now characterize the set $\calc$ of classical
states. A necessary and sufficient condition for classicality of a
state $\rho\in\caln$ is $Z\geq 0$, where $Z$ is given by
Eq.~\eqref{condC}. The characteristic polynomial of $Z$ reads
\begin{equation}
\det\left(x \idmat_3-Z\right)=x^3-\tr Z x^2+
\frac{(\tr Z)^2-\tr Z^2}{2} x-\det Z.
\end{equation}
Since $Z$ is real symmetric the three roots of the characteristic
polynomial are real. As in the previous section, Descartes' rule
of signs implies that the roots are positive, i.e.~$\rho$ is a
density matrix associated with a classical state, if and only if
the three conditions
\begin{equation}
\label{conditionsZ}
\tr Z\geq 0,\, (\tr Z)^2\geq\tr Z^2\, \textrm{ and } \det Z\geq 0
\end{equation}
are fulfilled. In terms of $\bfu$ and $W$ one has
\begin{eqnarray}
\tr Z&=&1-|\bfu|^2\label{zwu1}\\
\tr Z^2&=&\tr W^2-2\bra{\bfu}W\ket{\bfu}+|\bfu|^4\label{zwu2}\\
\det Z&=&\det \left(W-\ket{\bfu}\bra{\bfu}\right).\label{zwu3}
\end{eqnarray}
Using \eqref{limit0onu} and \eqref{zwu1} we see that condition
$\tr Z\geq 0$ is fulfilled by any density matrix. The two
remaining conditions on $Z$ do not depend on the basis in which
$\bfu$  and $W$ is expressed, thus we can write them in the basis
where $W$ is diagonal with eigenvalues $\mu_x,\mu_y,\mu_z$.
Using Eqs.~\eqref{zwu1}--\eqref{zwu2}, condition $(\tr Z)^2\geq\tr
Z^2$ is equivalent to
\begin{equation}
\label{condC1}
\sum_au_a^2(1-\mu_a)\leq \mu_x\mu_y+\mu_x\mu_z+\mu_y\mu_z.
\end{equation}
Condition $\det Z\geq 0$ becomes
\begin{equation}
\label{ellipsoidC}
\mu_y\mu_z u_x^2+\mu_x\mu_z u_y^2+\mu_x\mu_y u_z^2\leq\mu_x\mu_y\mu_z.
\end{equation}
A state $\rho$ belongs to $\calc$ if and only if it verifies
Eqs.~\eqref{limitonmu}--\eqref{limit0onu} and
\eqref{condC1}--\eqref{ellipsoidC}. A necessary condition for $Z$ to be positive
is that its diagonal elements $\mu_a-u_a^2$ are positive, which entails
positivity of the $\mu_a$ and thus $\mu_a\in[0,1]$.
If all $\mu_a$ differ from 0 and 1 then
\eqref{condC1} and \eqref{ellipsoidC} describe ellipsoids in
$u$-space, with axes lengths respectively given by $r_a$ and $r'_a$ with
\begin{equation}
r_a^2=\frac{\mu_x\mu_y+\mu_x\mu_z+\mu_y\mu_z}{1-\mu_a},\ \ \ \
{r'_a}^2=\mu_a
\end{equation}
Since all $\mu_a\in ]0,1[$, one has $r_a>r_a'$, thus
    \eqref{ellipsoidC} is more restrictive than \eqref{condC1}. It is
    also more restrictive than the equation of the sphere
    Eq.~\eqref{limit0onu} since $r_a'<1$. If $\mu_a=0$ or  $\mu_a=1$ for at least one value of $a$, one has
to consider all equations again.
Finally $\rho\in\calc$ if and only if the parameters $\mu_a$ and $u_a$ correspond to the following situations:
\begin{enumerate}
\item All $\mu_a\in ]0,1[$ and
\begin{equation}
\label{ensembleC}
\frac{u_x^2}{\mu_x}+\frac{u_y^2}{\mu_y}+\frac{u_z^2}{\mu_z}\leq 1,
\end{equation}
that is, $\mathbf{u}$ corresponds to a point inside an ellipsoid centered at $(0,0,0)$ with half-axes of length $\sqrt{\mu_a}$;
\item Exactly one of the $\mu_a$ is equal to 0, say $\mu_z=0$. Then from \eqref{ellipsoidC} one must have $u_z=0$ and from  \eqref{condC1}
\begin{equation}
\label{flatensembleC}
\frac{u_x^2}{\mu_x}+\frac{u_y^2}{\mu_y}\leq 1.
\end{equation}
This corresponds to the situation above flattened to 2 dimensions;
\item Two $\mu_a$ are zero, e.~g.~$\mu_y=\mu_z=0$. Then $\mu_x=1$
and from \eqref{condC1} one must have $u_y=u_z=0$, which leaves
the condition $|u_z|\leq 1$. Again this corresponds to the
situation \eqref{ensembleC}, flattened to 1 dimension.
\end{enumerate}

A point $\rho$ belongs to the boundary $\partial\calc$ of $\calc$
if one of the inequalities \eqref{conditionsZ} becomes an
equality. States with one or two $\mu_a$ equal to 0 always verify
$\det Z=0$ and thus lie on the boundary $\partial\calc$. When all
$\mu_a$ are in $]0,1[$, the condition that $\det Z=0$ is
equivalent to equality in \eqref{ensembleC}, which corresponds to
points $u_a$ which lie on the surface of the ellipsoid
\eqref{ensembleC}. Condition $\tr Z=0$ is equivalent to equality
in \eqref{limit0onu}, while condition $(\tr Z)^2=\tr Z^2$ is
equivalent to equality in \eqref{condC1}. Since the ellipsoid
\eqref{ensembleC} lies inside both the sphere \eqref{limit0onu}
and the ellipsoid \eqref{condC1}, the points corresponding to
either of these cases must lie on the surface of the ellipsoid
\eqref{ensembleC}. Therefore, points on the boundary
$\partial\calc$ correspond to case 1 above when \eqref{ensembleC}
becomes an equality, or to cases 2 or 3.

In the main case (equality in \eqref{ensembleC}) the surface is a two-parameter set of
ellipsoids that can be parametrized by $\mu_a\in]0,1[$ with
$\sum_a\mu_a=1$, e.~g.,
\begin{equation}
\label{parametrizationCmu}
\bfmu=\begin{pmatrix}
\sin^2\theta_1\sin^2\varphi_1 \\
\sin^2\theta_1\cos^2\varphi_1 \\
\cos^2\theta_1
\end{pmatrix},
\end{equation}
and
\begin{equation}
\label{parametrizationC}
\bfu=\begin{pmatrix}
\sin\theta_1\sin\varphi_1\cos\theta_2\cos\varphi_2 \\
\sin\theta_1\cos\varphi_1\cos\theta_2\sin\varphi_2 \\
\cos\theta_1\sin\theta_2.
\end{pmatrix}
\end{equation}
Again, the parametrization requires three more angles to take into account
the orthogonal matrix required to diagonalize $W$. If
$\alpha,\beta$ and $\gamma$ are the three Euler angles that parametrize the
orthogonal matrix $O$ then one has the
complete parametrization
\begin{eqnarray}
\label{paramuw}
\bfu&=&O(\alpha,\beta,\gamma)\begin{pmatrix}
\sin\theta_1\sin\varphi_1\cos\theta_2\cos\varphi_2 \\
\sin\theta_1\cos\varphi_1\cos\theta_2\sin\varphi_2 \\
\cos\theta_1\sin\theta_2
\end{pmatrix},\\
W&=&O\begin{pmatrix}
\sin^2\theta_1\sin^2\varphi_1&&0 \\
&\sin^2\theta_1\cos^2\varphi_1&\\
0&&\cos^2\theta_1\\
\end{pmatrix}O^{T}
\nonumber
\end{eqnarray}
with $O=O(\alpha,\beta,\gamma)$.
To summarize, the number of essential parameters for the set ${\cal
  C}$ (excluding rotations of the coordinate system) is 5, and for the
surface $\partial {\cal C}$ it is 4.

In order to obtain each classical matrix $\rho$ once and only once
we shall demand that $\mu_x\le\mu_y\le\mu_z$ which means that the
range in \eqref{parametrizationCmu}--\eqref{paramuw} has to be
restricted to $\theta_1\in ]0,\arctan (1/\cos( \varphi_1))]$ and
$\varphi_1\in]0,\pi/4]$ (see \cite{Byrd01}).

\section{Some examples}
\label{exC}
We first give an example of a non-classical state. In section
\ref{paramN} we saw that  the case $\mu_y=\mu_z=1$ corresponds to
state $\rho=\ket{1,0}\bra{1,0}$. According to \cite{Giraud10} this is
the most quantum spin--1 state. Its Majorana representation
corresponds to two points diametrically opposed on the Bloch sphere,
e.g.~north and south pole. 

Let us now consider the case where two of the $\mu_a$ vanish
(case 3 of the above section), say, $\mu_y=\mu_z=0$. Then
$\mu_x=1$ and Eq.~\eqref{ensembleN} implies that
$u_y=u_z=0$. Then $\rho\in\calc$ if and only if
$u_x=u\in[-1,1]$. In this case the ellipsoid is flattened to
a line. The state $\rho$ can be decomposed as
\begin{equation}
\rho=\frac{1-u}{2}\ket{\psi^{(-)}}\bra{\psi^{(-)}}+\frac{1+u}{2}\ket{\psi^{(+)}}\bra{\psi^{(+)}},
\end{equation} with \begin{equation}
\ket{\psi^{(\pm)}}=\frac{\ket{1,-1}\pm\sqrt{2}\ket{1,0}+\ket{1,1}}{2}.
\end{equation} The pure states $\ket{\psi^{(\pm)}}$ are
eigenvectors of $J_x$ corresponding to the eigenvalues $\pm
1$, i.~e., they are coherent states directed along or
opposite to the $x$-axis.
Since $u\in[-1,1]$, $\rho$ is a
classical mixture of two coherent states. It forms a
one-parameter family of classical states. Since the entire
family is inside $\partial\calc$, this represents a
one-dimensional line on the surface. This
indicates that the surface  $\partial\calc$ is not necessarily
strictly convex. Nevertheless, we now show that the surface $\partial\calc$
does not contain facets, that is,  the surface is not locally a
(7-dimensional) hyperplane.  

\begin{figure}
\includegraphics[scale=0.5]{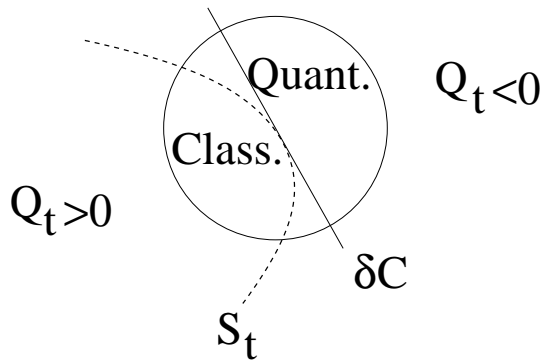}
\caption{Local geometry at a point on the surface $\partial\calc$ of the set of
  classical 
  states. }\label{fig2}
\end{figure}

For a state $\rho\in\caln$ and a three-dimensional real vector $\mathbf{t}$
with $|\mathbf{t}|=1$, we 
define
\begin{equation}
\label{Qt}
Q_{\mathbf{t}}=2\langle J_{\mathbf{t}}^2\rangle-\langle J_{\mathbf{t}}\rangle^2-1.
\end{equation}
As noted in \cite{Giraud08}, the classicality criterion $Z\geq 0$ is equivalent
to $Q_{\mathbf{t}}\geq 0$ for all $\mathbf{t}$. For fixed $\mathbf{t}$,
$Q_{\mathbf{t}}=0$  
defines a quadric surface $S_{\mathbf{t}}$ in the eight-dimensional space of
variables $\{u_a, W_{ab} \}$.  
Its equation can be rewritten as
$Q_{\mathbf{t}}=\sum_{a,b}(W_{ab}-u_au_b)t_at_b=0$, with $t_a$, $a=x,y,z$,
fixed. 
Each surface $S_{\mathbf{t}}$ separates the space of all states $\caln$ into
two subsets.  The subset of 
states with $Q_{\mathbf{t}}<0$ contains only genuinely quantum states, 
as they violate the condition $Q_{\mathbf{t}}\geq 0$ for at least one
$\mathbf{t}$. 
The subset of states with $Q_{\mathbf{t}}>0$ is convex: indeed, a linear
change of variables 
with a new variable $X_1=\sum_a u_a t_a$ yields $Q_{\mathbf{t}}=X_1^2+$
linear terms. Now let $M$ be a point on $\partial\calc$ and suppose there exists a sphere
$B_{\epsilon}$ with radius $\epsilon$, 
centered on $M$, such that inside  $B_{\epsilon}$ the surface
$\partial\calc$ is a piece of a hyperplane 
(see Fig.~\ref{fig2}). By definition of $\calc$, the sphere is split 
into two equal halves, one containing only quantum states and the other one
containing only classical states. But since  $S_{\mathbf{t}}$ is not a flat
surface, 
part of the states in the latter half-sphere must lie on the
subset on states with $Q_{\mathbf{t}}<0$ (see Fig.~\ref{fig2}), which entails a contradiction.

\begin{figure}
\includegraphics[scale=0.5]{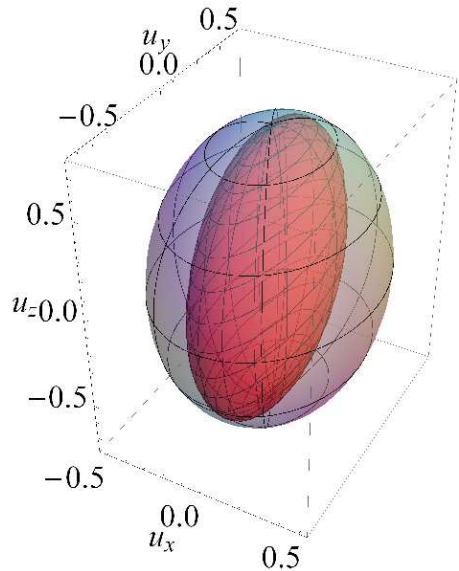}
\caption{(Color online) Boundaries $\partial\caln$ of the set of
  physical density matrices (outer ellipsoid) and
  $\partial\calc$ of the classical states (inner ellipsoid)
  for $\mu_x=0.05$, $\mu_y=0.4$,
and $\mu_z=0.55$ in terms of the dimensionless components $u_i$ of the
  vector ${\bf u}$
  defined in eq.(\ref{canonrhoj1}). The two ellipsoids do not touch in general, and their axes coincide.}\label{fig1}
\end{figure}

Another interesting example is the thermal state
\begin{equation}
\rho=e^{-\beta H}/{\rm tr}e^{-\beta H}
\label{thermal}
\end{equation}
of a system with Hamiltonian $H=J_z^2$ and inverse temperature $\beta=1/k_BT$,
with $k_B$ Boltzmann's constant. For temperature $T=0$, the thermal state
is the ground state $|1,0\rangle$, which is the most quantum state
possible.  For
$T\to\infty$ on the other hand, $\rho$ approaches the identity matrix
and is therefore classical. The transition temperature to classicality
can be found exactly from the boundary $\partial\calc$.  The parametrization of
\eqref{thermal} gives ${\bf u}={\bf 0}$ and
\begin{equation}
W=\left(\begin{array}{ccc}
\frac{e^\beta}{2+e^\beta}&0&0\\
0&\frac{e^\beta}{2+e^\beta}&0\\
0&0&\frac{2-e^\beta}{2+e^\beta}\end{array}\right)\ .
\end{equation}
The inequality in (\ref{ensembleC}) is always
satisfied. The condition that $\mu_a\in [0,1]$ reduces to $0\leq e^\beta\leq 2$.
Therefore, $\rho$ is classical if and only if $\beta\le \ln 2$.

As a last example, consider a state with $\mu_x=0.05$, $\mu_y=0.4$,
and $\mu_z=0.55$. One can then specify the boundaries $\partial\calc$
and $\partial\caln$ solely in terms of the $u_a$.  Fig.\ref{fig1} shows
that $\partial\calc$ is indeed an ellipsoid inside the ellipsoid
given by $\partial\caln$.

\section{Conclusions}
To summarize, we have found an explicit representation of the set
$\calc$ of classical spin--1 states, Eq.~(\ref{ensembleC}) defined
as the convex hull of spin--1 SU(2) coherent states.  The set
$\calc$ consists of a family of ellipsoids.  The surface of the
set contains straight lines, thus this allows the existence of
linear families of genuinely quantum states with exactly the same
quantumness.
 Our results allow to visualize the set of classical states and to determine
 analytically 
under what conditions a density matrix that depends on one or several
parameters becomes genuinely ``quantum''.

{\em Acknowledgments:} DB thanks Otfried G\"uhne for discussions. This work
has been supported by the   
Sonderforschungsbereich TR 12 of the Deutsche
Forschungsgemeinschaft and by the GDRI-471 of the CNRS.

\section*{Appendix}
Let $E_0(\kappa)$ be the lowest eigenvalue of the
parameter-dependent Hermitian matrix $ H(\kappa)$ and $\psi_0$ its
corresponding eigenvector. At values of $\kappa$ such that the
ground level is non-degenerate the second derivative of
$E_0(\kappa)$ is  non-positive.  This follows, e.g., from the
identity $E''_0=-2\langle\psi_0| T_0|\psi_0\rangle$ with $T_0$
denoting a manifestly positive operator,
\begin{eqnarray}
 T_0=
\frac{\partial  H}{\partial\kappa}%
 Q_0\left( H-E_0\right)^{-1}Q_0%
\frac{\partial H}{\partial\kappa},\nonumber\\%
Q_0=\hat1-\left|\psi_0\left\rangle\right\langle\psi_0\right|.
\end{eqnarray}

Assume now that at $\kappa=0$ the ground state is non-degenerate
and besides, the first derivative $E'_0(0)=0$. Then for all
positive $\kappa$ the ground state energy will be monotonically
decreasing (or at best non-growing) function of $\kappa$. Indeed,
if we first assume that $E_0(\kappa)$ is not degenerate for all
$k\geq 0$ then we have $E'_0(\kappa)=\int_0^\kappa E''_0(x)dx\le
0$ for $\kappa\ge 0$. The result remains true even if there is
level crossing at some
 $\kappa=\kappa_1$ since then for $\kappa
>\kappa_1$ we can write
$E'_0(\kappa)=E'_0(\kappa_1+0^{+})+\int_{\kappa_1+0^{+}}^\kappa E''_0(x) dx$.
The integral is negative from the same argument as above, and
$E'_0(\kappa_1+0^{+})$ is the slope immediately after the crossing, which must be
smaller than the slope immediately before the crossing, which we know to be negative.

\bibliography{../../mybibs_bt}
\end{document}